\documentclass[a4paper,fleqn]{cas-dc}

\usepackage[numbers,sort,compress]{natbib}

\def\tsc#1{\csdef{#1}{\textsc{\lowercase{#1}}\xspace}}
\tsc{WGM}
\tsc{QE}
\tsc{EP}
\tsc{PMS}
\tsc{BEC}
\tsc{DE}

\usepackage{bm}
\usepackage{booktabs}
\usepackage{color}
\usepackage{soul}

\begin{document}
\let\WriteBookmarks\relax
\def\floatpagepagefraction{1}
\def\textpagefraction{.001}
\let\printorcid\relax

\shorttitle{Target detection of integrated camera \& radar}

\shortauthors{C. Zhu, Z. Zhao, et~al.}

\title [mode = title]{Robust Target Detection of Intelligent Integrated Optical Camera and mmWave Radar System}                      



%
\author[1,2]{Chen Zhu}



\ead{zhuc@zju.edu.cn}

\affiliation[1]{organization={College of Engineering, Zhejiang University},
    city={Hangzhou},
    country={China}}

\affiliation[2]{organization={National Engineering Elite Program, Zhejiang University},
    city={Hangzhou},
    country={China}}
    
\author[3]{Zhouxiang Zhao}
\ead{zhouxiangzhao@zju.edu.cn}
\affiliation[3]{organization={College of Information Science and Electronic Engineering, Zhejiang University},
    city={Hangzhou},
    country={China}}
    
\author[3]{Zejing Shan}
\ead{shanzejing@zju.edu.cn}

\author[4]{Lijie Yang}
\ead{yanglijie@zhejianglab.com}
\affiliation[4]{organization={Zhejiang Lab},
    city={Hangzhou},
    country={China}}

\author[5]{Sijie Ji}
\ead{sijieji@hku.hk}
\affiliation[5]{organization={The University of Hong Kong},
    city={Hong Kong},
    country={China}}

\author[3]{Zhaohui Yang}
\ead{yang_zhaohui@zju.edu.cn}
\cormark[1]

\author[3]{Zhaoyang Zhang}
\ead{ning_ming@zju.edu.cn}

\cortext[cor1]{Corresponding author}



\begin{abstract}
Target detection is pivotal for modern urban computing applications. While image-based techniques are widely adopted, they falter under challenging environmental conditions such as adverse weather, poor lighting, and occlusion. To improve the target detection performance under complex real-world scenarios, this paper proposes an intelligent integrated optical camera and millimeter-wave (mmWave) radar system. Utilizing both physical knowledge and data-driven methods, a long-term robust radar-camera fusion algorithm is proposed to solve the heterogeneous data fusion problem for detection improvement. For the occlusion scenarios, the proposed algorithm can effectively detect occluded targets with the help of memory through performing long-term detection. For dark scenarios with low-light conditions, the proposed algorithm can effectively mark the target in the dark picture as well as provide rough stickman imaging. The above two innovative functions of the hybrid optical camera and mmWave radar system are tested in real-world scenarios. The results demonstrate the robustness and significant enhancement in the target detection performance of our integrated system.
\end{abstract}




\maketitle

\section{Introduction}
Target detection serves as a critical underpinning for contemporary urban computing systems, facilitating a myriad of applications ranging from traffic control to surveillance and autonomous vehicles \cite{1703855,6678247,9314219,10028264,8704212,8425968,9624519}. The development of modern communication technology \cite{xu2023edge,10032275,zhao2023semantic,chen2023big} also requires highly reliable target detection techniques. While traditional image-based detection techniques have proven effective in a variety of contexts, they exhibit limitations when faced with complex real-world conditions such as inclement weather (e.g., fog, smog, heavy rain) or suboptimal lighting conditions. In contrast, millimeter-wave (mmWave) radar technology has gained prominence as a robust alternative, owing to its unique capabilities \cite{8732419,9824264}. Specifically, mmWave radar can penetrate airborne particulates and maintain functionality across a wide range of environmental conditions, including adverse weather and low-light scenarios. Moreover, unlike image-based approaches that are contingent upon the visual characteristics of the target, mmWave radar offers a level of invariance to these factors.
Nevertheless, mmWave radar is not without its challenges. One of the primary limitations lies in the quality of the point cloud data it generates for target detection. These data are often sparse and contaminated with noise, attributable to factors such as low angular resolution, specular reflections, and multi-path effects. Consequently, achieving high-accuracy target detection solely through mmWave radar remains challenging. The growing demand of urban computing urges a robust and accurate target detection solution. 

The complementary properties of optical cameras and mmWave radar have led to many research endeavors\cite{joshi2016review,zhou2019optical,iepure2021novel,7163620,7289083} to design fusion methods for robust target detection. Traditional fusion methods~\cite{cho2014multi}, usually use a Kalman Filter to fuse the sensor data, which may oversimplify the radar-based detection to a point target detection and require a hand-crafted fusion strategy. With the development of deep learning technology, more fusion methods have been proposed. Liu et al. \cite{liu2022fusing} proposed a feature-level fusion method that enhances 3D object detection, as validated on the NuScenes dataset. Lin et al. \cite{lin20223d} developed an algorithm for multi-target tracking, which utilizes the fusion of mmWave radar and camera data to estimate 3D bounding boxes of traffic objects accurately. Deng et al. \cite{deng2022global} introduced the Global-Local Feature Enhancement Network (GLE-Net), a deep fusion detector that excels in challenging weather and lighting conditions, showing superior average precision. Li et al. \cite{li2022pedestrian} presented a feature fusion network based on attention mechanisms, effectively enabling real-time pedestrian detection. Although these methods report excellent performance, they are limited by specific datasets that are collected by simulation or controlled environment and have yet to be fully realized in practical applications. On the other hand, various fusion systems have been proposed. Zhang et al. \cite{zhang2019extending} unveiled a radar-camera fusion system that addresses the complexities of fusing data from heterogeneous sensors. Their system employs a fusion extended Kalman filter, achieving remarkable distance and angular accuracy in real-time applications. Batra et al. \cite{batra2022fusion} focused on indoor target recognition, offering a solution for extracting geometric attributes by fusing optical and mmWave synthetic aperture radar images.
Shuai et al. \cite{shuai2021millieye} introduced a lightweight mmWave radar and camera fusion system that can adapt to new scenes by requiring a small amount of labeled data from the new scene. Differing from the above systems, this paper introduces a novel fusion method that aims for robust target detection even in complex environments.

In particular, this paper introduces a novel integrated system that synergizes optical camera and mmWave radar technologies. Leveraging both physical knowledge and data-driven methods, our approach aims to provide a robust solution for target detection under low-lighting conditions, accommodate varying detection scenarios with multiple targets, and effectively handle mutual target occlusion. In addition, the system offers advanced visualization capabilities, providing intuitive and informative detection results that enhance situational awareness and decision-making across a range of applications. Through this integrated system design, we aim to bridge existing gaps and contribute to the advancement of the ISCC domain, particularly in the realm of target detection and recognition.

\begin{table*}
    \centering
    \caption{Pros and cons of optical camera and mmWave radar}
    \begin{tabular}{cccccccc}
        \hline
         &  Resolution&  Color &  All-Weather &  Cost&  Detection& Penetration\\
           &  &   Information&  Capability&  & Range&Through Obstacles\\
         \hline
         Optical Camera&  High&  Yes&  No&  Cheap&  Short& No\\ \hline
         mmWave Radar&  Low&  No&  Yes&  Cheap&  Long& Yes\\ \hline
    \end{tabular}
    
    \label{tab1}
\end{table*}

\section{System Design}
In this section, we first introduce the background properties of the optical camera and mmWave radar that support us to design the system and then we delve into the comprehensive design of our fusion-based detection system, which combines optical and mmWave radar technologies for enhanced target detection performance.

\subsection{Background and Design Intuition}
The integration of high-precision optical and mmWave imaging modalities is predicated on their complementary electromagnetic characteristics, engendering enhanced imaging capabilities. Optical imaging is proficient in capturing salient features such as the location, appearance, morphology, and material properties of unobstructed scatterers in well-lit conditions \cite{9562196}. Conversely, mmWave sensing is particularly efficacious in low-illumination scenarios, facilitating the detection of occluded targets through its transmission capabilities \cite{9266605}.

Optical imaging, characterized by its considerably higher frequency compared to mmWaves, is subject to significant energy loss in the form of reflection, scattering, refraction, and transmission in the vast majority of cases when visible light signals are employed. Consequently, optical imaging predominantly focuses on elucidating the direct path between the observer and the target object. This optical imaging approach affords two primary advantages. Firstly, the received optical imaging signal inherently contains only direct path information, obviating the necessity for intricate signal processing algorithms. The optical camera can directly capture optical information, thereby providing a rudimentary perception of the environment. Secondly, the optical camera is equipped to discern information pertaining to the target object's material characteristics, discernable through color, brightness, and darkness cues. However, the optical camera grapples with challenges in acquiring distance and speed information when confronted with occluded data. It is crucial to note that the optical camera operates passively, its perceptual quality contingent upon ambient light intensity. Notably, it cannot function autonomously in conditions such as darkness and smoke.

In contrast, electromagnetic sensing encounters certain limitations in the form of sparsity and constrained range. As the distance from the sensor increases, the number of data points returned diminishes significantly. Consequently, distant targets may yield scant or no data points, rendering them individually undetectable. Concurrently, the image data yielded by the camera exhibits high density, which proves advantageous for semantic understanding tasks like object detection and target segmentation. Leveraging its high resolution, the camera excels at identifying distant targets, albeit with relatively less precision in distance measurement.

In the context of mmWave imaging, the electromagnetic attributes derived from optical imaging serve as prior knowledge, effectively mitigating spatial uncertainty and thereby augmenting computational efficiency and accuracy \cite{tong2023multi}. Conversely, mmWave's electromagnetic properties supplement and enrich the data acquired through optical imaging, contributing to a more holistic environmental understanding. Table 1 shows the pros and cons of optical camera and mmWave radar. Guided by these principles, we design an integrated optical camera and mmWave radar system. The optical imaging module initially furnishes a preliminary scene model, encapsulating parameters such as the position, orientation, and surface attributes of walls and visible targets. This data, amalgamated with prior knowledge of various scatterer types, informs a generative model to produce an initial environmental perception.

Moreover, the system exploits temporal properties to further enhance performance. As the sensing modalities iteratively inform each other, the imaging results undergo refinement by incrementally increasing resolution and focusing on previously ambiguous regions or elusive targets. By collaborative use of optical camera and mmWave radar, leveraging their respective advantages, utilizing prior knowledge, and optimizing imaging algorithms, we aim for a robust and accurate target detection system.

\subsection{System Overview}

\begin{figure*}[t]
\centering
\includegraphics[width=6.5in]{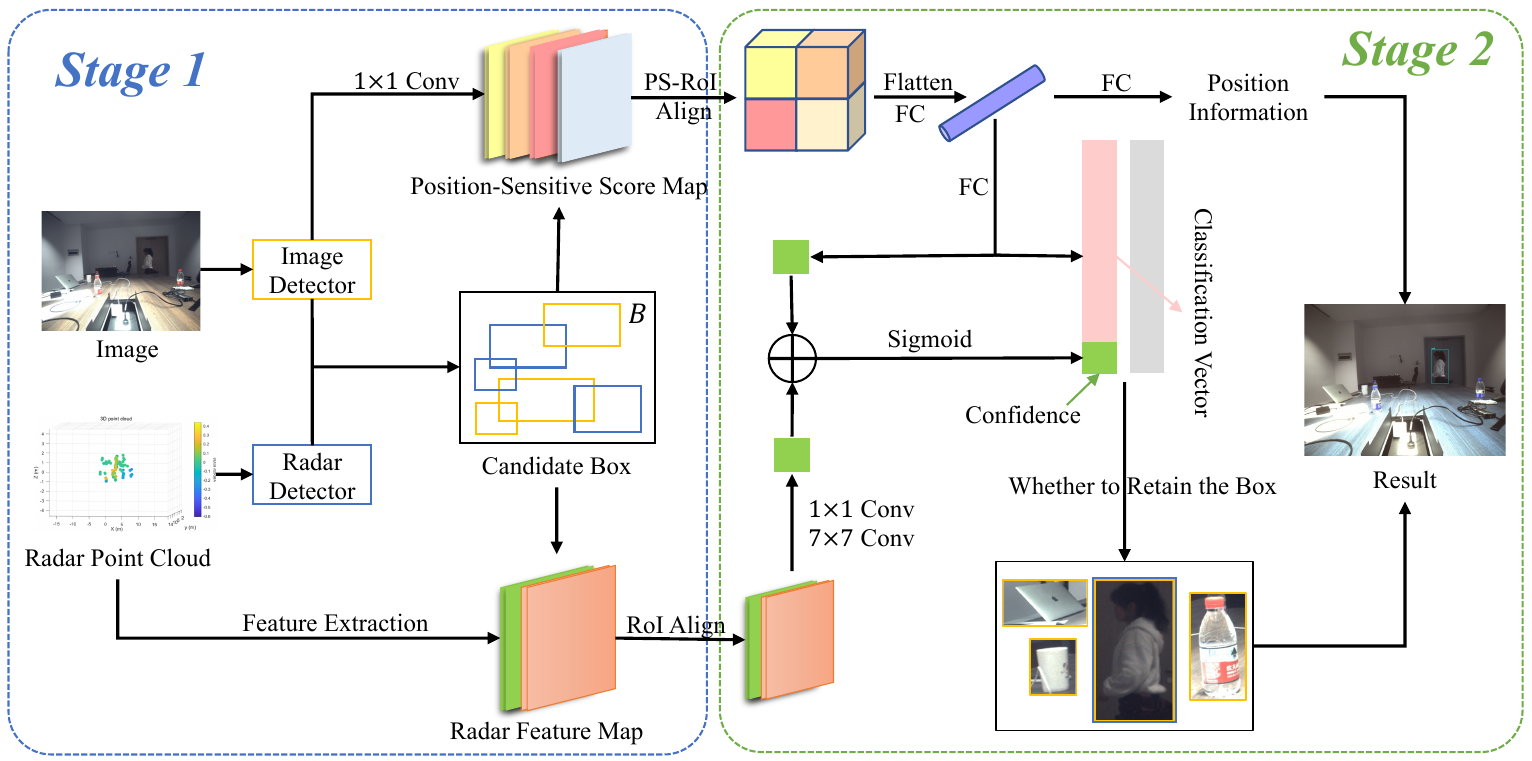}
\vspace{-.5em}
\caption{Framework of the system model.}
\vspace{-.5em}
\label{fig2}
\end{figure*}

Our system model follows a two-stage detection paradigm, aiming to fuse optical and mmWave radar data effectively. The overall framework is depicted in Fig. \ref{fig2}. In contrast to one-stage image detection algorithms, the two-stage approach incorporates a detection box refinement module to enhance accuracy. This refinement module benefits from mmWave radar data, which provides insights into object occupancy, facilitating more precise discrimination between foreground and background. Even in extreme conditions where traditional optical image-based detectors falter, the radar-based module can generate reliable detection candidates, ensuring uninterrupted system operation.

In the first stage, candidate boxes are aggregated from both the image detector (I) and the radar detector (R) to form a set denoted as $B=\{B^I,B^R\}=\{b_k\}_{k=1}^K$, where $K=|B|$ represents the total number of region-of-interest (RoI). Both the image and radar branches extract global features from their respective data sources, producing multi-modal feature maps $G^I$ and $G^R$. Subsequently, local features for each RoI are obtained through cropping operations on these feature maps: $L^I,L^R=\operatorname{Cropping}(G^I,G^R;B)$. These sets comprise local feature maps for both modalities: $L^I=\{l_m^I\}_{m=1}^M$ and $L^R=\{l_n^R\}_{n=1}^N$.

In the second stage, a fused refinement header is implemented to predict new positions and confidence scores for each candidate box. These refined bounding boxes: $b_k^\prime=\operatorname{Refinement}(l_m^I,l_n^R,b_k)$, offer improved reliability compared to the original bounding boxes ($b_k$) due to the fusion of information from both modalities. The decision to retain or discard a bounding box is made based on the newly calculated confidence score, subject to a predefined threshold.

The integration of image and radar detectors, along with the refinement modules, is achieved through a loosely coupled architecture. This design flexibility allows the system to accommodate various image detectors, including the single-stage detector YOLO \cite{redmon2018yolov3} and the single-shot detector (SSD) \cite{liu2016ssd}. During training, image-related modules can be individually fine-tuned using large-scale image datasets, with parameters subsequently fixed. This design approach reduces the need for extensive multimodal data during operation, demonstrating the model's strong generalization capabilities.

\subsection{Image Detector}
The image detector within our model leverages the widely adopted YOLO v3 algorithm, which is built upon a convolutional neural network (CNN). The modularity and ease of replacement of this module are significant advantages of our model. 

The workflow of the image detection module begins with the input image, denoted as $I$, which is processed through the feature extractor $F_\text{body}$. This feature extractor typically comprises a convolutional layer, an activation layer, and a pooling layer. The output of this process is an internal feature map, represented as $f=F_\text{body}(I)$.

Subsequently, the extracted feature map $f$ undergoes further processing through the network $F_\text{head}$. This processing step generates a series of bounding boxes, denoted as $B$. In essence, the entire workflow of the image detector can be summarized as follows: $B=F_\text{head}\left(F_\text{body}(I)\right)$.

In the subsequent refinement operation, the bounding boxes generated by the image detector, referred to as $B$, are aggregated with the bounding boxes produced by the radar detector. This aggregation forms the comprehensive set of candidate frames after being filtered by confidence scores. Additionally, the feature map $f$ is subjected to a $1 \times 1$ convolutional layer to create a location-sensitive score map. This score map serves as a crucial input in the second stage of the refinement process, significantly reducing computational demands and optimizing performance.

\begin{figure*}[t]
\centering
\includegraphics[width=6.5in]{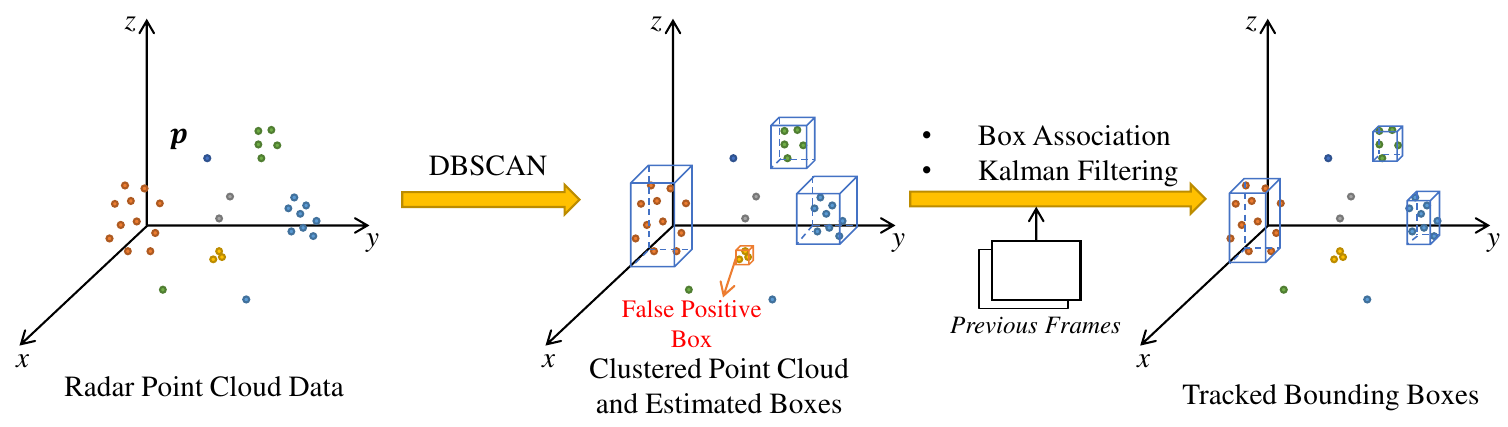}
\vspace{-.5em}
\caption{Radar detector module.}
\vspace{-.5em}
\label{fig3}
\end{figure*}

\subsection{Radar Detector}

The radar detection module, depicted in Fig. \ref{fig3}, plays a crucial role in our model. This module processes the point cloud data acquired by the radar sensor, with each point $p$ in the point cloud map represented as a four-dimensional vector: 
\begin{equation}\label{eq1}
p_i \triangleq (x_i,y_i,z_i,v_i) \in \mathbb{R}^4.
\end{equation}
These dimensions correspond to three-dimensional coordinates ($x$, $y$, $z$) and radial velocity ($v$).

To address the presence of clutter and noise in the point cloud data, we employ Density-Based Spatial Clustering of Applications with Noise (DBSCAN) \cite{ester1996density}. DBSCAN effectively filters out points generated by clutter, as it groups points from foreground objects into clusters while scattering points from clutter at low densities. Unlike K-means, DBSCAN doesn't require prior knowledge of the number of clusters, making it suitable for object detection tasks with varying numbers of objects. The distance between two points, $d(i,j)$, is defined as follows:
\begin{multline}\label{eq2}\vspace{-.5em}
    d(i,j) \triangleq \alpha_x \left(x_i-x_j\right)^2+\alpha_y \left(y_i-y_j\right)^2+\\
    \alpha_z \left(z_i-z_j\right)^2+\alpha_v \left(v_i-v_j\right)^2,
\end{multline}
where $\alpha=\left[\alpha_x,\alpha_y,\alpha_z,\alpha_v\right]$ is a weight vector that balances the contribution of each element. This distance metric is utilized in DBSCAN for density connectivity checking. Importantly, velocity information is incorporated in the clustering process, aiding in distinguishing nearby objects with different velocities, such as when two individuals pass by each other.

After DBSCAN, each point is labeled with a cluster index or an outlier flag. Outliers are filtered out, and the center position and velocity along the $z$-axis of each cluster are estimated by averaging the corresponding values for all points within that cluster. Additionally, for each cluster, the outermost points belonging to it are identified, and these points are used to approximate the size of a 3D bounding box. The 3D bounding box for each cluster is defined as:
\begin{equation}\label{eq3}\vspace{-.5em}
    z \triangleq (x,y,z,v_z,w,h,t) \in \mathbb{R}^7,
\end{equation}
where $w$, $h$, and $t$ represent the width, height, and thickness of the 3D bounding box, respectively.

To ensure temporal consistency and eliminate false alarms in subsequent frames, our model leverages the Hungarian algorithm \cite{kuhn1955hungarian} to associate bounding boxes across frames. The Euclidean distance between the centers of any two frames serves as the matching metric. To further reduce jitter in associated frames from neighboring frames, a Kalman filter is employed. At frame $N-1$, the Kalman filter maintains a state variable $s_{i,N-1}$ for the $i$th bounding box:
\begin{equation}\label{eq4}\vspace{-.5em}
    s_{i,N-1} \triangleq (x,y,z,v_x,v_y,v_z,w,h,t) \in \mathbb{R}^9.
\end{equation}

At frame $N$, a continuous velocity model is used to predict a new state vector, $s_{i,N-1}^\prime$, for each bounding box. This state vector is then corrected based on observations from frame $N$:
\begin{equation}\label{eq5}\vspace{-.5em}
    s_{i,N-1} = s_{i,N-1}^\prime + K\left(z_{i,N}-H s_{i,N}^\prime\right),
\end{equation}
where $K\in \mathbb{R}^{9\times 7}$ is the Kalman gain matrix and $H\in \mathbb{R}^{7\times 9}$ is the observation model matrix. It's noteworthy that the state vector's length is 9, while the observation vector's length is 7 because mmWave radar does not provide velocities in the $x$ and $y$ directions.

If a bounding box cannot be associated with a new one in the next frame, prediction of a new state vector will continue using the constant velocity model. If $T_\text{max}$ successive frames fail to establish an association, the object is assumed to have disappeared, and prediction ceases.

To facilitate fusion between the two sensors, which necessitates a uniform coordinate system and timestamp, the 3D bounding boxes generated in the radar coordinate system are projected into the 2D image. This process involves slicing the 3D bounding box on the $z$-axis to obtain cross-sections, which are then projected into the 2D image. The projection of each point is computed using the equation:
\begin{equation}\label{eq6}\vspace{-.5em}
\left[\begin{array}{l}
u \\
v \\
1
\end{array}\right]=\frac{1}{z} K T\left[\begin{array}{l}
x \\
y \\
z \\
1
\end{array}\right],
\end{equation}
where $K$ represents a $3\times 3$ internal camera matrix, and $T$ is a $3\times 4$ external camera matrix. $(x, y, z)$ denote the 3D position coordinates in the radar coordinate system, while $(u, v)$ are the projected pixel coordinates in the image coordinate system. As the relative positions of the two sensors remain fixed, both $K$ and $T$ can be computed offline.

\subsection{Refinement Module}
To harness the radar's occupancy detection capabilities while preserving the robust image feature extraction performance derived from extensive image datasets, our model incorporates a fusion-capable refinement head. This head comprises four key components: a RoI-based Convolutional Neural Network (R-CNN) subnetwork, perceptual fusion, integration, and a multi-frame detection module, as illustrated in Fig. \ref{fig2}.

The R-CNN subnetwork utilizes knowledge learned from a large volume of image data to perform regression and classification tasks on the detected frames. It capitalizes on the wealth of information present in the image dataset to enhance the accuracy of object detection. The perceptual fusion module consolidates confidence scores from both sensing modalities, i.e., optical and radar. This fusion of confidence scores allows for a more comprehensive assessment of object detection, taking advantage of the strengths of each modality. The integration module further leverages the intelligence obtained from image-based detectors to enhance prediction reliability. By combining information from both modalities, it ensures more robust object detection. The multi-frame detection component relies on tracking detected frames across multiple frames, mitigating issues arising from occlusion. This approach addresses the problem of missing detection frames, which can occur when objects are obscured from view.

The fusion and integration modules do not rely on image features, making them insensitive to variations in appearance. Consequently, these components can be trained with a limited amount of multi-modal labeled data, demonstrating their adaptability and robustness. The subsequent sections detail the operations and feature extraction steps for each RoI within this framework.

\subsubsection{Feature Extraction of Image and Radar Data}
The image feature extraction process begins by isolating the internal feature maps from the image detection module. These feature maps are then subjected to a $1\times 1$ convolutional layer to generate a position-sensitive (PS) score map, which comprises 490 channels. The PS score map is subsequently cropped using a PS-RoI Align layer, with cropping being determined by the position of the bounding box. This operation results in the creation of a $7\times 7\times 10$ feature map for each Region of Interest (RoI). For detailed implementation specifics of the PS-RoI Align layer, readers are referred to the work by Li et al. \cite{li2017light}.

Before extracting features from radar data, a preprocessing step is employed to address the issue of sparse radar point clouds. The radar point cloud is transformed into a two-dimensional image, leveraging the robust feature extraction capabilities of CNNs. This conversion process involves three key steps:
\begin{enumerate}
    \item Projecting the point cloud into 2D image coordinates;
    \item Calculating the 2D histograms of the projected point cloud on the following three channels: 
    \begin{itemize}
        \item Number of points on the $z$-axis
        \item Average depth
        \item Average velocity
    \end{itemize}
    \item Normalizing the values on each channel to the range of $[0,1]$.
\end{enumerate}
Following these preprocessing steps, a 3-channel heat map is generated. This heat map is then passed through a 3-layer CNN to extract the occupancy feature map, which encodes the probability of the target's presence at each location. Subsequently, the RoI alignment layer \cite{he2017mask} is applied to crop the feature map, resulting in RoI radar features, each of size $7\times 7\times 10$. These extracted features are essential for subsequent stages of the refinement module and aid in improving detection accuracy.

\subsubsection{R-CNN Subnetwork}
R-CNN refers to a type of CNN architecture that is specifically designed to work with RoI within an input image. This approach is commonly used in object detection tasks, where the goal is to identify and locate objects of interest within an image.

The RoI typically represents a specific part or region of the input image where the network focuses its attention. Instead of processing the entire image at once, the R-CNN selectively processes only the regions that are likely to contain relevant information.

R-CNN family includes R-CNN itself, Fast R-CNN, and Faster R-CNN. These models use a two-stage approach. In the first stage, potential RoI are generated using a region proposal network (RPN). In the second stage, these proposed regions are fed into a CNN for further analysis and classification.

In the R-CNN subnetwork, confidence scores are refined by incorporating information from the radar branches, enhancing their reliability. The network begins by flattening the feature maps derived from the images. Subsequently, a fully connected layer is applied, followed by two additional fully connected layers dedicated to bounding box regression and classification. For each RoI image feature map, the process involves flattening, applying a fully connected (FC) layer with 256 channels, and concatenating two fully connected layers. This results in the generation of a 4-dimensional vector for bounding box regression and a $(C+1)$-dimensional vector for classification, where $C$ represents the number of object classes.

\subsubsection{Perceptual Fusion}
Despite the sparse and low-resolution nature of the radar point cloud, it provides strong indications of object presence, making it a crucial complement to the confidence scores obtained from the R-CNN subnetwork. To achieve this fusion, each radar feature within the RoI undergoes processing with two convolutional layers using kernel sizes of $7\times 7$ and $1\times 1$. This process abstracts a global representation that also incorporates the radar confidence scores. Subsequently, the confidence scores from both modalities are summed and passed through a sigmoid activation function, resulting in the final confidence score.

\subsubsection{Integration}
The R-CNN subnetwork and the perceptual fusion module jointly produce a $(C+1)$-dimensional vector with the same dimensions as the output from the image detector. To effectively merge these two $(C+1)$-dimensional vectors and derive a more reliable classification decision, the model incorporates a learning-based integration module. This integration module is composed of two fully connected layers and an activation layer. Its output is instrumental in distinguishing between foreground and background objects. By setting an appropriate threshold, users can determine the list of bounding boxes to be retained.

In essence, the integration module evaluates the consistency of classification results between the two input vectors. A higher output confidence score is indicative of greater inter-category variance, signifying a more dependable classification result. It's essential to note that this module skips the bounding boxes generated by the radar tracker, as the image-based target detector lacks corresponding detection results for integration.

\subsubsection{Multi-frame Detection}
In real-world scenarios, multi-target detection often faces the challenge of mutual occlusion, where neither the image detector nor the radar tracker can provide a reliable occlusion decision. To address this issue, a multi-frame detection algorithm is proposed, leveraging the observation that objects, such as people, do not suddenly vanish from the scene. If the number of detected individuals in the current frame is fewer than in the previous frame, the multi-frame detection module conducts a comparison of detection frame positions between consecutive frames while considering their relative velocities. This analysis enables the identification of lost detection frames by matching distance and speed criteria within the obscured region. Additionally, to account for scenarios where a target may be initially occluded, then move away before disappearing, resulting in an extended period of lost detection, the algorithm defines a threshold of $N$ consecutive frames of absence to confirm target disappearance. This multi-frame detection algorithm effectively mitigates occlusion challenges, enhancing the accuracy and robustness of multi-target detection while providing more precise human target localization.

\subsubsection{Visualization}
In challenging scenarios like low light or dense fog, where discerning target detection results can be difficult, a position complementation technique based on detection frames is introduced. When lighting conditions are inadequate, this method augments target information onto the original image using the positional data from the detection frame. This augmentation enhances the intuitiveness and comprehensibility of detection results. By applying this approach, more comprehensive and precise target information can be acquired, facilitating target detection and information recovery under varying lighting conditions. This innovation offers fresh insights and methodologies for enhancing the resilience and adaptability of future target detection techniques.

\subsection{Training Strategy and Loss Function}
The model separates the training of image-related and radar-related modules, allowing the system to learn on various large image datasets thus having good robustness and generalization capabilities, while jointly optimizing the fusion of radar and image data using small multi-modal datasets. The whole training of the model is divided into three steps: 
\begin{enumerate}
    \item Training the target detector;
    \item Fixing the parameters trained by the image detector to train the R-CNN subnetwork;
    \item Fine-tuning the radar-related parts on the multi-modal dataset.
\end{enumerate}
The first step can be further separated as dataset preparation, exploratory data analysis, data preprocessing, data segmentation, machine learning algorithm modeling, and machine learning task selection. Since the first two steps involve only image data, they can be performed on large image datasets.

The third stage of training requires that the minimization objective function of the neural network should contain two parts. Firstly the integration module decides whether the bounding box of each RoI should be preserved or not, which is a regression problem, so the model uses the focal loss function[22] which is defined as
\begin{equation}\label{eq7}\vspace{-.5em}
    L_{\text{Focal},i}=\left\{
    \begin{array}{cl}
    -\alpha\left(1-p_i\right)^\gamma \log p_i &,y_i=1 \\
    -(1-\alpha) p_i^\gamma \log \left(1-p_i\right) &,y_i=0
    \end{array}\right.,
\end{equation}
where $y_i\in [0,1]$ is the labelling about retaining or discarding the detection frames, $p_i$ is the prediction confidence score of the integration module, $\alpha$ is a factor to balance the positive and negative samples, and $\gamma$ is a conditioning factor during the training process. Since the candidate detection frames from the radar tracker do not involve the integration module, this loss function is computed only for the candidate detection frames from the image detector.

Furthermore, in order to force the fusion module to mimic the behavior of the binary classifiers in order to generate a reliable confidence score about whether the proposed RoI is a positive instance or not, the binary cross entropy (BCE) loss is used[22], which is given by the following equation
\begin{equation}\label{eq8}\vspace{-.5em}
    L_{\text{BCE},i}=-y_i\log q_i,
\end{equation}
where $q_i$ is the confidence score predicted by the fusion module.

For training stability, only the losses of positive and negative samples are computed with intersection over union (IoU) greater than 0.7 and less than 0.3, respectively[22]. To balance multi-task training, the final loss is the weighted sum of the above two terms
\begin{equation}\label{eq9}\vspace{-.5em}
    L=\sum_{i\in \text{pos}\cup \text{neg}} \left[\mathbb{1} (i\in \text{img})\cdot L_{\text{Focal},i}+\lambda L_{\text{BCE},i}\right].
\end{equation}
where  pos is the positive samples, neg is the negative samples, img is the collection of bounding boxes from the image detector, and $\lambda$ is the weighting factor.

\begin{figure}[t]
\centering
\includegraphics[width=3in]{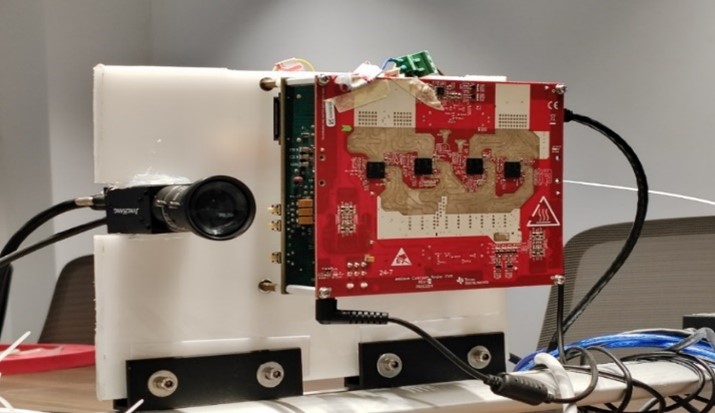}
\vspace{-.5em}
\caption{Radar-camera sensor suite.}
\vspace{-.5em}
\label{fig4}
\end{figure}

\begin{figure*}[t]
\centering
\includegraphics[width=3in]{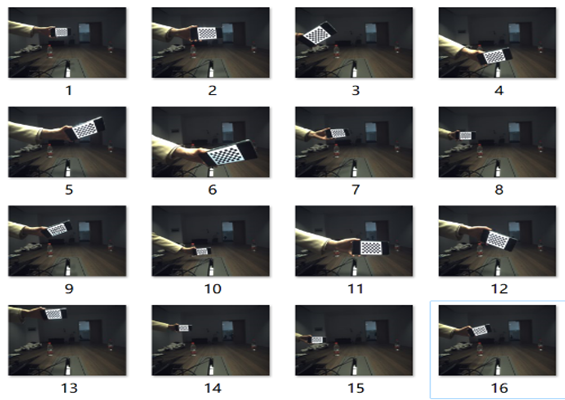}
\includegraphics[width=3.5in]{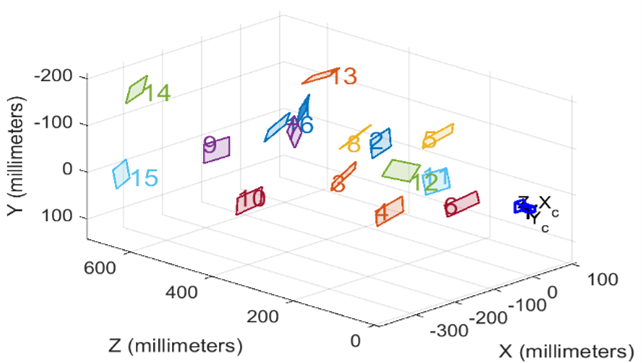}
\vspace{-.5em}
\caption{Calibration data acquisition and post-calibration attitude restoration.}
\vspace{-.5em}
\label{figs5}
\end{figure*}

\begin{figure}[t]
\centering
\includegraphics[width=3in]{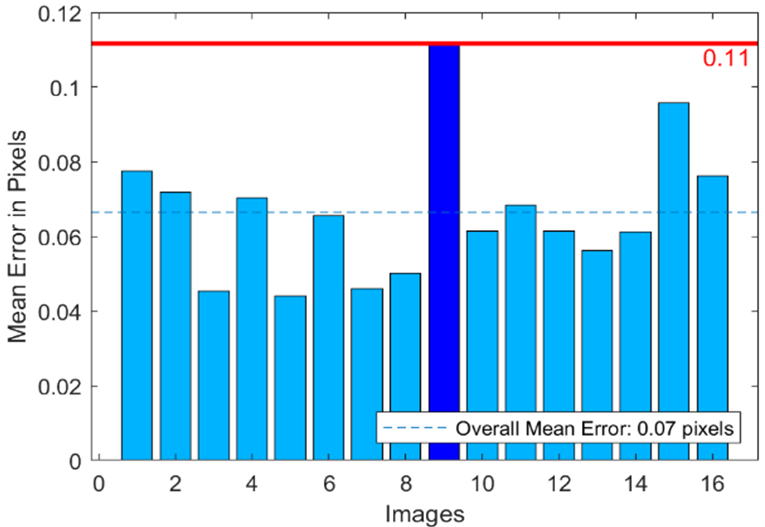}
\vspace{-.5em}
\caption{Camera calibration error.}
\vspace{-.5em}
\label{fig6}
\end{figure}

\begin{table*}
    \centering
    \caption{Main Parameters of the Industrial Camera}
    \begin{tabular}{cccccccc}
        \hline
         &  Model Number&  Monitoring &  Aperture &  Sizes&  Distortion& Optical Back\\
           &  &  Range [mm] & Range &  & &Focus[mm]\\
         \hline
         Industrial Camera&  VM1220MP5&  12&  F2.0-C&  1/1.8”&  <0.1\%& 12.5\\ \hline
    \end{tabular}
    
    \label{tab2}
\end{table*}

\begin{table*}
    \centering
    \caption{Main Parameters of the mmWave Radar}
    \begin{tabular}{cccccccc}
        \hline
         &  Start\_freq[GHz] &  Slope[MHz/us] &  Idle\_time[us] &   Sample\_freq[ksps] & Rx\_gain[db] \\
         \hline
         mmWave Radar&  77 &  79 &  5 &   8000 & 48\\ \hline
    \end{tabular}
    
    \label{tab3}
\end{table*}

\begin{table}[ht]
\centering
\caption{Camera Calibration Parameters}
\begin{tabular}{cc}
    \toprule
    \textbf{Parameter}                                  & \textbf{Value} \\
    \midrule
    Image Size                                          & $[1536,2048]$   \\
    Radial Distortion                                   & $[-0.09635,0.08026]$         \\
    Intrinsic Matrix                                    & $[1208.2,0,1038.8;$       \\ 
                                                        & $0,1210.4,763.4;0,0,1]$      \\
    Tangential Distortion                               & $[0,0]$     \\
    \bottomrule
\end{tabular}
\label{tb1}
\end{table}

\section{System Implementation and Analysis}
We implemented the target detection system that synergistically integrates a JHEM Gigabit network industrial camera with a TEXAS INSTRUMENT four-chip mmWave cascade radar. Table 2 shows the main parameters of the industrial camera, and Table 3 shows the main parameters of the mmWave radar. In order to collect data for our experiments, we carried out a series of trials comprising 118 instances of single-class human detections. The purpose of these experiments was to obtain data in diverse conditions, encompassing various lighting scenarios and multiple targets.
A random walk in front of the sensors was arranged to simulate realistic scenarios for data collection. The sensor suite employed for data acquisition is illustrated in Fig. \ref{fig4}. 
To ensure the quality and comprehensiveness of the data, we configured the radar with the following settings: a frequency range spanning from 77 to 81 GHz, a maximum bandwidth of 4 GHz, a detection range of up to 10 meters, a distance resolution of 5 cm, and the capability to capture target velocities up to 26 km/h. These parameter settings were instrumental in acquiring high-quality data that served as the empirical foundation for our subsequent experiments and analysis.

\subsection{Calibration}
In this paper, a four-chip mmWave cascade radar is selected, which will have array error, and the array error mainly includes the amplitude and phase inconsistency between the channels of each array element, the mutual coupling between the array elements and so on. In order to correct the array error, this experiment adopts the pass channel calibration method to eliminate the amplitude and phase errors.

In this experiment, we utilized MATLAB's Camera Calibrator tool to execute camera calibration, adhering to the Zhang Youzheng calibration methodology as delineated in reference \cite{791289}. The calibration process was initiated by capturing images of a $6\times 9$ checkerboard grid, strategically positioned at varying distances, orientations, and angles relative to the camera.
Fig. \ref{figs5} showcases the assortment of images acquired under these diversified conditions, along with the grid points detected and mapped in the camera's coordinate system. After calibration, we achieved an average error metric of 0.07 pixels, a testament to the precision of the calibration procedure. The intrinsic camera parameters obtained from the calibration are detailed in Table \ref{tb1},
and the calibration errors are visualized in Fig. \ref{fig6}. These calibration steps allowed us to acquire precise camera intrinsic parameters, ensuring more accurate results in subsequent tasks, including but not limited to, camera orientation estimation and target tracking applications.

\subsection{Radar Data Processing}
The raw radar data collected during the experiment undergo an essential processing pipeline to yield the radar point cloud map, which serves as input for the model. This processing pipeline comprises several key modules, including ADC data calibration, distance FFT, Doppler FFT, CFAR, and DOA.

The first stage involves the ADC data calibration module, which is instrumental in rectifying nonlinear errors inherent in the analog-to-digital conversion process. This module also mitigates offset and gain errors, thereby enhancing the fidelity of the subsequent processing steps.
Following this, the distance FFT module utilizes Fourier Transform techniques to convert time-domain signals into their frequency-domain counterparts, facilitating the generation of the point cloud map based on distance information. In parallel, the Doppler FFT module focuses on capturing velocity-related data, facilitating the construction of a velocity-centric point cloud map.
The CFAR module, a specialized signal processing technique, is employed for the precise detection and localization of targets within the radar echo signals. Concluding the pipeline, the DOA module leverages the aggregated information to calculate key parameters such as the target's position and velocity, culminating in the final radar point cloud map.

\begin{figure*}[t]
\centering
\includegraphics[width=6.7in]{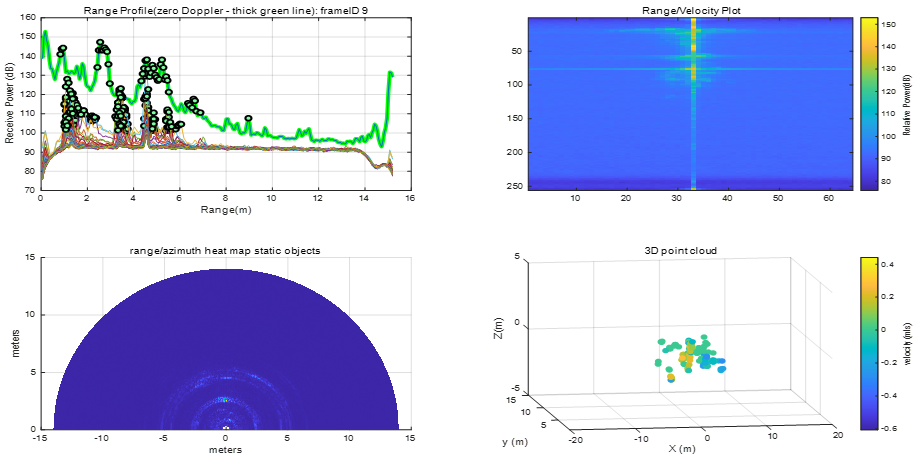}
\vspace{-.5em}
\caption{Sample of radar raw data processing results.}
\vspace{-.5em}
\label{fig7}
\end{figure*}

As illustrated in Fig. \ref{fig7}, these processing pipelines effectively transform the raw radar data into a coherent and informative radar point cloud map, laying a crucial foundation for subsequent data analysis and processing.

\begin{figure*}[t]
\centering
\includegraphics[height=2in]{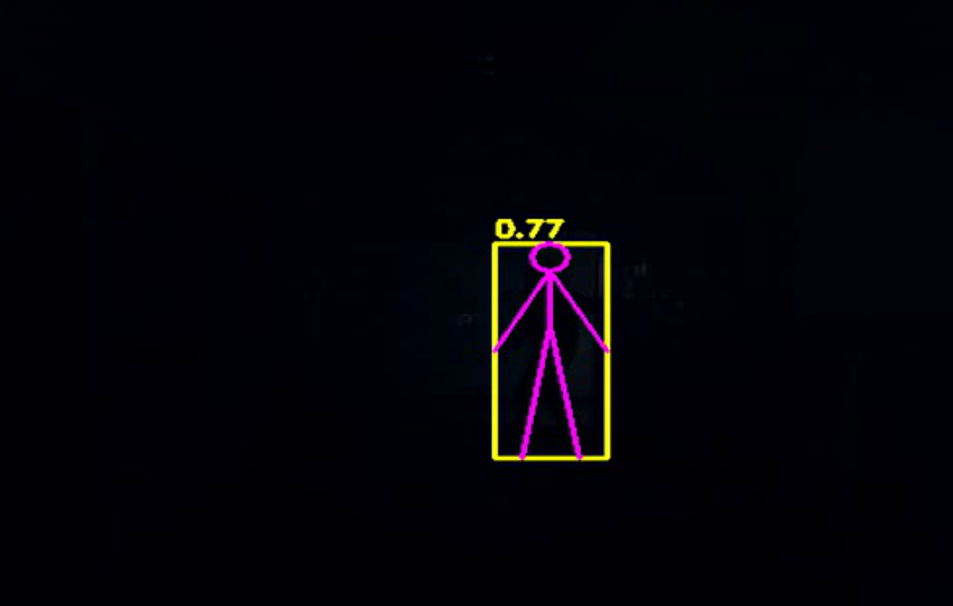}
\includegraphics[height=2in]{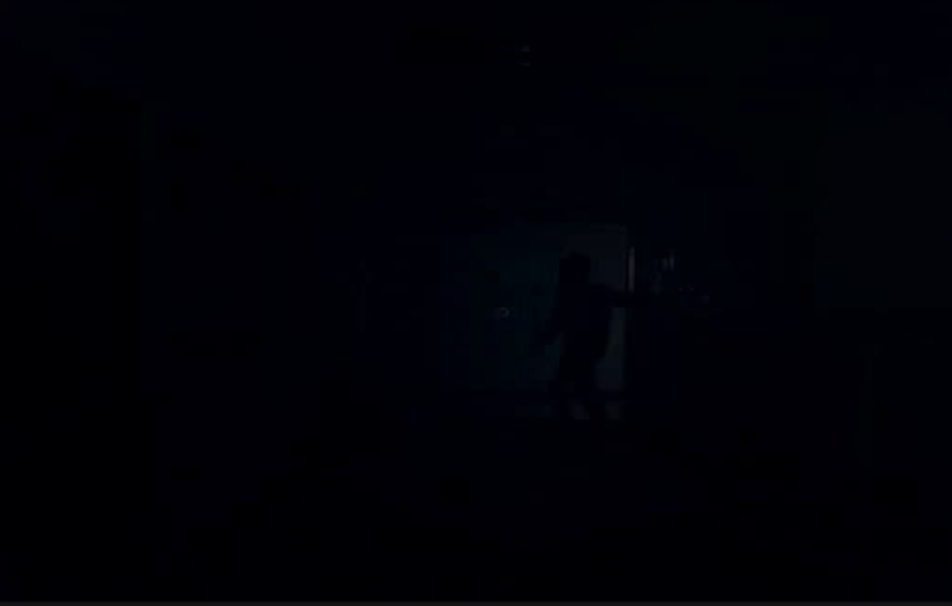}
\vspace{-.5em}
\caption{Target detection result in dark environments with (left) and without (right) radar fusion techniques.}
\vspace{-.5em}
\label{figs8}
\end{figure*}

\subsection{Experimental Results Analysis}
\subsubsection{Fusion vs Image-only}
Fig. \ref{figs8} presents a compelling comparison between the results obtained from the camera and radar fusion model and those from the image detection-only model in a low-light environment. In the left figure, we observe the output of the fusion model, which combines radar and camera data, while the right figure showcases the output of the YOLO v3 image detector operating in isolation.

Notably, in low-light conditions, the conventional image-based detector struggles to accurately detect all targets due to the limited availability of visual information. Conversely, the fusion model, enriched with radar data, continues to exhibit more precise target detection. This stark contrast underscores the fusion model's remarkable effectiveness in challenging low-light environments.

Traditional image-based approaches often falter, yielding suboptimal detection rates in practical challenging adverse conditions, making the fusion model proposed in this paper exceptionally valuable. The empirical results substantiate the fusion model's robustness and accuracy in target detection, particularly in low-light settings. These findings not only validate the methodological soundness of our approach but also offer compelling evidence for its practical utility in enhancing target detection performance.

\subsubsection{Multi-targets detection}
\begin{figure*}[t]
\centering
\includegraphics[width=.925\linewidth]{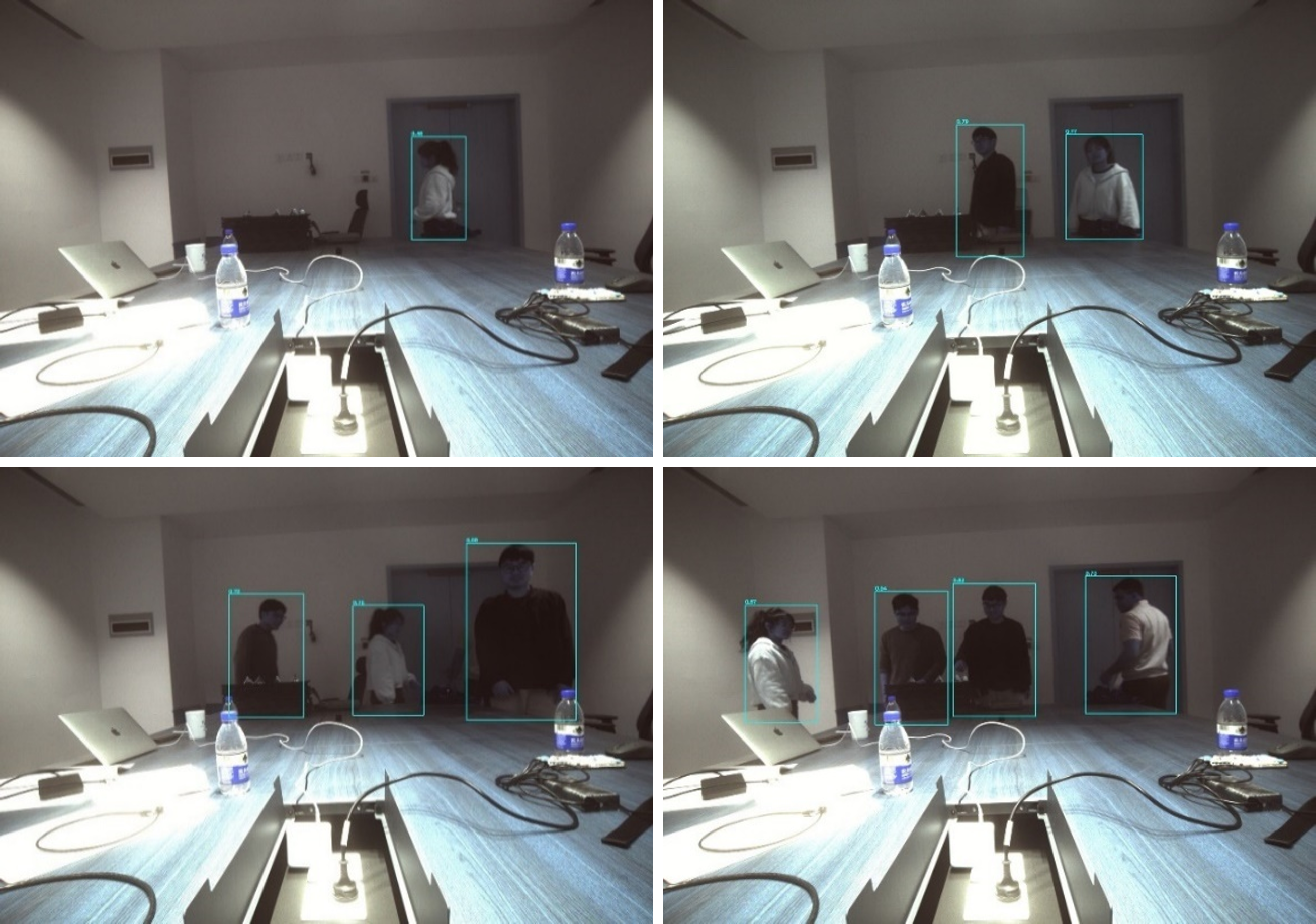}
\vspace{-.5em}
\caption{Multi-target detection effect.}
\vspace{-.5em}
\label{figs9}
\end{figure*}

The detection results of the model under varying numbers of detected targets are visually presented in Fig. \ref{figs9}. Notably, the model demonstrates its capability to accurately detect targets across scenarios with both a small number of detected targets and a large number of detected targets. This remarkable performance underscores the model's robust generalization ability, illustrating its aptitude for adapting to diverse scenarios and addressing varying detection challenges.
The ability to maintain high detection accuracy across varying targets is indicative of the model's robust learning and generalization capabilities. Such versatility renders the model well-suited for  a wide range of detection tasks, extending its utility to various real-world scenarios.

\begin{figure*}[t]
\centering
\includegraphics[height=2in]{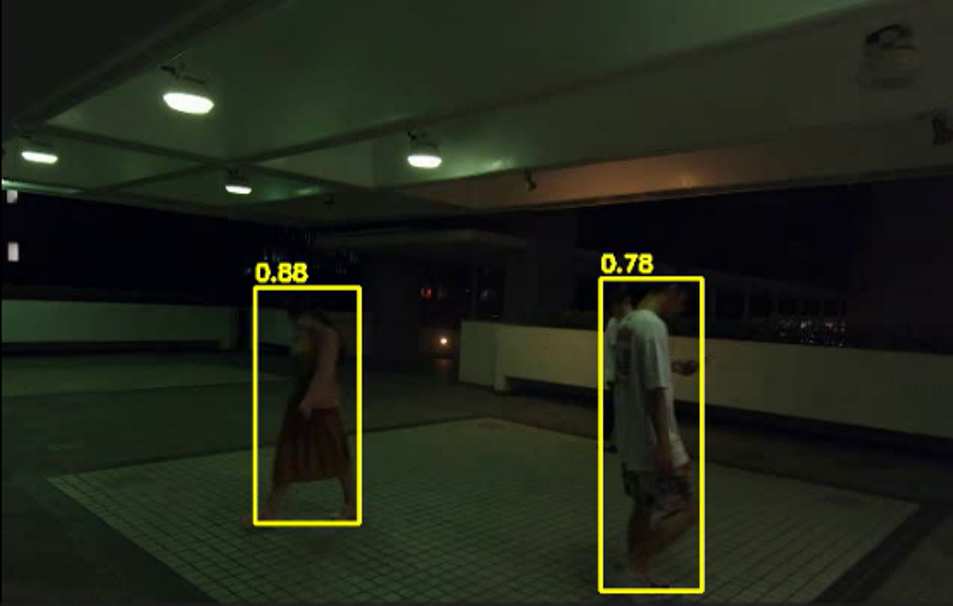}
\includegraphics[height=2in]{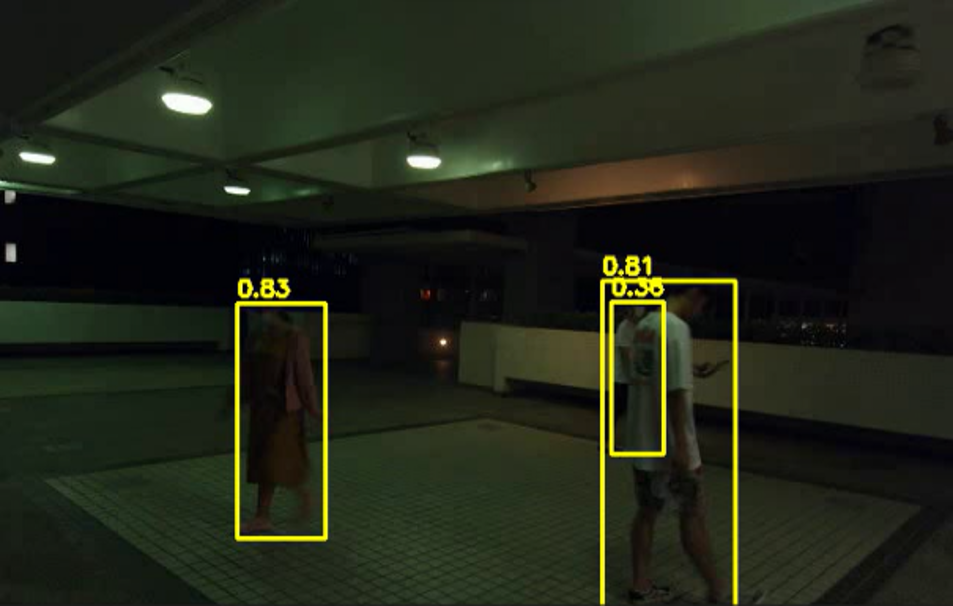}
\vspace{-.5em}
\caption{Comparison of the effect before and after adding multi-frame detection module.}
\vspace{-.5em}
\label{figs10}
\end{figure*}

\subsubsection{Multi-frame detection}
A comparison of the results before and after adding the multi-frame detection module is shown in Fig. \ref{figs10}. The comparison illustrates that the model using the multi-frame detection module excels in detecting occluded targets, even in complex scenarios involving mutual target occlusion. These experimental results underscore the advantages of the multi-frame detection module in enhancing the model's robustness and addressing occlusion challenges effectively. By leveraging motion information across multiple frames, the multi-frame detection module significantly improves both the accuracy and stability of target tracking and detection. This highlights the pivotal role and value of the multi-frame detection module in the proposed fusion-based target detection system.

\section{Conclusion}
In conclusion, this paper provides a comprehensive analysis of the imaging characteristics of optical and mmWave technologies, laying the theoretical foundation for the development of a high-performance joint optical and mmWave target detection system. Moreover, it introduces a multi-frame joint detection approach based on tracking information, enhancing the accuracy of RoI imaging in the radar-camera fusion algorithm. This method effectively addresses the issue of target loss due to occlusion during the tracking process, ultimately improving the accuracy of the mmWave radar and optical image fusion-based target detector.

Furthermore, the paper presents visualizations of detection results in extreme environments with low lighting conditions, enhancing the interpretability of the outcomes. Experimental results demonstrate the model's strong generalization capability and its ability to enhance detection efficiency across various lighting conditions and different numbers of detected targets.

Despite the results achieved in this paper, there are some directions for further research. Firstly, advanced algorithms that utilize the depth information provided by the mmWave radar can be considered to be explored in the fusion system to extend the target detection from 2D to 3D. Secondly, the visualization part of the results in this paper also deserves further optimization, e.g. the use of multi-array radar for position recognition in dark environments can be considered.

\section*{Declaration of competing interest}
The authors declare that they have no known competing financial interests or personal relationships that could have appeared to influence the work reported in this paper.

\section*{Acknowledgments}
This work is supported by Zhejiang Lab Program under grant K2023QA0AL02, and Zhejiang Science and Technology Program under grant 2023C01021.



\bibliography{cas-refs}

\begin{thebibliography}{10}

\bibitem{1703855}
I.~Bekkerman and J.~Tabrikian.
\newblock Target detection and localization using mimo radars and sonars.
\newblock {\em IEEE Transactions on Signal Processing}, 54(10):3873--3883, 2006.

\bibitem{6678247}
Nasser~M. Nasrabadi.
\newblock Hyperspectral target detection : An overview of current and future challenges.
\newblock {\em IEEE Signal Processing Magazine}, 31(1):34--44, 2014.

\bibitem{9314219}
Yimian Dai, Yiquan Wu, Fei Zhou, and Kobus Barnard.
\newblock Attentional local contrast networks for infrared small target detection.
\newblock {\em IEEE Transactions on Geoscience and Remote Sensing}, 59(11):9813--9824, 2021.

\bibitem{10028264}
Xiaolong Chen, Jian Guan, Zhigao Wang, Hai Zhang, and Guoqing Wang.
\newblock Marine targets detection for scanning radar images based on radar- yolonet.
\newblock In {\em 2021 CIE International Conference on Radar (Radar)}, pages 1256--1260, 2021.

\bibitem{8704212}
Xiaohan Yu, Xiaolong Chen, Yong Huang, Lin Zhang, Jian Guan, and You He.
\newblock Radar moving target detection in clutter background via adaptive dual-threshold sparse fourier transform.
\newblock {\em IEEE Access}, 7:58200--58211, 2019.

\bibitem{8425968}
Xiaolong Chen, Baoxin Chen, Jian Guan, Yong Huang, and You He.
\newblock Space-range-doppler focus-based low-observable moving target detection using frequency diverse array mimo radar.
\newblock {\em IEEE Access}, 6:43892--43904, 2018.

\bibitem{9624519}
Zhuoqun Liu, Yingjie Deng, Feng Ma, Jinming Du, Chao Xiong, Moufa Hu, Luping Zhang, and Xuhuan Ji.
\newblock Target detection and tracking algorithm based on improved mask rcnn and lmb.
\newblock In {\em 2021 International Conference on Control, Automation and Information Sciences (ICCAIS)}, pages 1037--1041, 2021.

\bibitem{xu2023edge}
Wei Xu, Zhaohui Yang, Derrick Wing~Kwan Ng, Marco Levorato, Yonina~C Eldar, and M{\'e}rouane Debbah.
\newblock Edge learning for b5g networks with distributed signal processing: Semantic communication, edge computing, and wireless sensing.
\newblock {\em IEEE journal of selected topics in signal processing}, 17(1):9--39, 2023.

\bibitem{10032275}
Zhaohui Yang, Mingzhe Chen, Zhaoyang Zhang, and Chongwen Huang.
\newblock Energy efficient semantic communication over wireless networks with rate splitting.
\newblock {\em IEEE Journal on Selected Areas in Communications}, 41(5):1484--1495, 2023.

\bibitem{zhao2023semantic}
Zhouxiang Zhao, Zhaohui Yang, Quoc-Viet Pham, Qianqian Yang, and Zhaoyang Zhang.
\newblock Semantic communication with probability graph: A joint communication and computation design.
\newblock {\em arXiv preprint arXiv:2310.00015}, 2023.

\bibitem{chen2023big}
Zirui Chen, Zhaoyang Zhang, and Zhaohui Yang.
\newblock Big ai models for 6g wireless networks: Opportunities, challenges, and research directions.
\newblock {\em arXiv preprint arXiv:2308.06250}, 2023.

\bibitem{8732419}
Theodore~S. Rappaport, Yunchou Xing, Ojas Kanhere, Shihao Ju, Arjuna Madanayake, Soumyajit Mandal, Ahmed Alkhateeb, and Georgios~C. Trichopoulos.
\newblock Wireless communications and applications above 100 ghz: Opportunities and challenges for 6g and beyond.
\newblock {\em IEEE Access}, 7:78729--78757, 2019.

\bibitem{9824264}
Ziwei Wan, Zhen Gao, Shufeng Tan, and Liang Fang.
\newblock Joint channel estimation and radar sensing for uav networks with mmwave massive mimo.
\newblock In {\em 2022 International Wireless Communications and Mobile Computing (IWCMC)}, pages 44--49, 2022.

\bibitem{joshi2016review}
Neha Joshi, Matthias Baumann, Andrea Ehammer, Rasmus Fensholt, Kenneth Grogan, Patrick Hostert, Martin~Rudbeck Jepsen, Tobias Kuemmerle, Patrick Meyfroidt, Edward~TA Mitchard, et~al.
\newblock A review of the application of optical and radar remote sensing data fusion to land use mapping and monitoring.
\newblock {\em Remote Sensing}, 8(1):70, 2016.

\bibitem{zhou2019optical}
Yejian Zhou, Lei Zhang, Yunhe Cao, and Yan Huang.
\newblock Optical-and-radar image fusion for dynamic estimation of spin satellites.
\newblock {\em IEEE Transactions on Image Processing}, 29:2963--2976, 2019.

\bibitem{iepure2021novel}
Bogdan Iepure and Aldo~W Morales.
\newblock A novel tracking algorithm using thermal and optical cameras fused with mmwave radar sensor data.
\newblock {\em IEEE Transactions on Consumer Electronics}, 67(4):372--382, 2021.

\bibitem{7163620}
Sicong Liu, Fang Yang, Wenbo Ding, and Jian Song.
\newblock Double kill: Compressive-sensing-based narrow-band interference and impulsive noise mitigation for vehicular communications.
\newblock {\em IEEE Transactions on Vehicular Technology}, 65(7):5099--5109, 2016.

\bibitem{7289083}
Sicong Liu, Fang Yang, Wenbo Ding, and Jian Song.
\newblock A priori aided compressive sensing approach for impulsive noise reconstruction.
\newblock In {\em 2015 International Wireless Communications and Mobile Computing Conference (IWCMC)}, pages 205--209, 2015.

\bibitem{cho2014multi}
Hyunggi Cho, Young-Woo Seo, BVK~Vijaya Kumar, and Ragunathan~Raj Rajkumar.
\newblock A multi-sensor fusion system for moving object detection and tracking in urban driving environments.
\newblock In {\em 2014 IEEE International Conference on Robotics and Automation (ICRA)}, pages 1836--1843. IEEE, 2014.

\bibitem{liu2022fusing}
Yangyang Liu, Shuo Chang, Zhiqing Wei, Kezhong Zhang, and Zhiyong Feng.
\newblock Fusing mmwave radar with camera for 3-d detection in autonomous driving.
\newblock {\em IEEE Internet of Things Journal}, 9(20):20408--20421, 2022.

\bibitem{lin20223d}
Zihao Lin and Jianming Hu.
\newblock 3d multi-object tracking based on radar-camera fusion.
\newblock In {\em 2022 IEEE 25th International Conference on Intelligent Transportation Systems (ITSC)}, pages 2502--2507. IEEE, 2022.

\bibitem{deng2022global}
Kaikai Deng, Dong Zhao, Qiaoyue Han, Zihan Zhang, Shuyue Wang, and Huadong Ma.
\newblock Global-local feature enhancement network for robust object detection using mmwave radar and camera.
\newblock In {\em ICASSP 2022-2022 IEEE International Conference on Acoustics, Speech and Signal Processing (ICASSP)}, pages 4708--4712. IEEE, 2022.

\bibitem{li2022pedestrian}
Hao Li, Ruofeng Liu, Shuai Wang, Wenchao Jiang, and Chris~Xiaoxuan Lu.
\newblock Pedestrian liveness detection based on mmwave radar and camera fusion.
\newblock In {\em 2022 19th Annual IEEE International Conference on Sensing, Communication, and Networking (SECON)}, pages 262--270. IEEE, 2022.

\bibitem{zhang2019extending}
Renyuan Zhang and Siyang Cao.
\newblock Extending reliability of mmwave radar tracking and detection via fusion with camera.
\newblock {\em IEEE Access}, 7:137065--137079, 2019.

\bibitem{batra2022fusion}
Aman Batra, Tobias Hark, Jonas Schorlemer, Nils Pohl, Ilona Rolfes, Michael Wiemeler, Diana G{\"o}hringer, Thomas Kaiser, and Jan Barowski.
\newblock Fusion of optical and millimeter wave sar sensing for object recognition in indoor environment.
\newblock In {\em 2022 Fifth International Workshop on Mobile Terahertz Systems (IWMTS)}, pages 1--5. IEEE, 2022.

\bibitem{shuai2021millieye}
Xian Shuai, Yulin Shen, Yi~Tang, Shuyao Shi, Luping Ji, and Guoliang Xing.
\newblock millieye: A lightweight mmwave radar and camera fusion system for robust object detection.
\newblock In {\em Proceedings of the International Conference on Internet-of-Things Design and Implementation}, pages 145--157, 2021.

\bibitem{9562196}
Neelima Devulapalli, Vicente Matus, Elizabeth Eso, Zabih Ghassemlooy, and Rafael Perez-Jimenez.
\newblock Lane-cross detection using optical camera-based road-to-vehicle communications.
\newblock In {\em 2021 17th International Symposium on Wireless Communication Systems (ISWCS)}, pages 1--5, 2021.

\bibitem{9266605}
Mohammad~Hossein Moghaddam, Sina~Rezaei Aghdam, Alessio Filippi, and Thomas Eriksson.
\newblock Statistical study of hardware impairments effect on mmwave 77 ghz fmcw automotive radar.
\newblock In {\em 2020 IEEE Radar Conference (RadarConf20)}, pages 1--6, 2020.

\bibitem{tong2023multi}
Xin Tong, Zhaoyang Zhang, and Zhaohui Yang.
\newblock Multi-view sensing for wireless communications: Architectures, designs, and opportunities.
\newblock {\em IEEE Communications Magazine}, 2023.

\bibitem{redmon2018yolov3}
Joseph Redmon and Ali Farhadi.
\newblock Yolov3: An incremental improvement.
\newblock {\em arXiv preprint arXiv:1804.02767}, 2018.

\bibitem{liu2016ssd}
Wei Liu, Dragomir Anguelov, Dumitru Erhan, Christian Szegedy, Scott Reed, Cheng-Yang Fu, and Alexander~C Berg.
\newblock Ssd: Single shot multibox detector.
\newblock In {\em Computer Vision--ECCV 2016: 14th European Conference, Amsterdam, The Netherlands, October 11--14, 2016, Proceedings, Part I 14}, pages 21--37. Springer, 2016.

\bibitem{ester1996density}
Martin Ester, Hans-Peter Kriegel, J{\"o}rg Sander, Xiaowei Xu, et~al.
\newblock A density-based algorithm for discovering clusters in large spatial databases with noise.
\newblock In {\em kdd}, volume~96, pages 226--231, 1996.

\bibitem{kuhn1955hungarian}
Harold~W Kuhn.
\newblock The hungarian method for the assignment problem.
\newblock {\em Naval research logistics quarterly}, 2(1-2):83--97, 1955.

\bibitem{li2017light}
Zeming Li, Chao Peng, Gang Yu, Xiangyu Zhang, Yangdong Deng, and Jian Sun.
\newblock Light-head r-cnn: In defense of two-stage object detector.
\newblock {\em arXiv preprint arXiv:1711.07264}, 2017.

\bibitem{he2017mask}
Kaiming He, Georgia Gkioxari, Piotr Doll{\'a}r, and Ross Girshick.
\newblock Mask r-cnn.
\newblock In {\em Proceedings of the IEEE international conference on computer vision}, pages 2961--2969, 2017.

\bibitem{791289}
Zhengyou Zhang.
\newblock Flexible camera calibration by viewing a plane from unknown orientations.
\newblock In {\em Proceedings of the Seventh IEEE International Conference on Computer Vision}, volume~1, pages 666--673 vol.1, 1999.

\end{thebibliography}


\bio{}
\quad \textbf{Chen Zhu} received his M.Eng. degree from Zhejiang University of Technology, Hangzhou, China in 2010. He is currently engaged in teaching and research at the College of Engineering, Zhejiang University. His research interests include image processing, target recognition, cloud-edge collaboration and IoT security.
\endbio

\bio{}
\quad \textbf{Zhouxiang Zhao} received his bachelor’s degree from College of Information Science and Electronic Engineering, Zhejiang University, Hangzhou, China in 2023. He is currently pursuing master’s degree in Zhejiang University. His research interests include semantic communication, and integrated sensing and communication.
\endbio

\bio{}
\quad \textbf{Zejing Shan} received her bachelor's degree from Zhejiang University, Hangzhou, China in 2023. She is currently pursuing master's degree in Tsinghua University, Beijing, China. Her research interests include digital video stabilization and semantic communication.
\endbio

\bio{}
\quad \textbf{Lijie Yang} was born in Zhejiang, China, in 1991. He received the B.S. degree and Ph.D. degree in electronic and information engineering from Zhejiang University, Hangzhou, China, in 2013 and 2019, respectively. He is now an assistant researcher in Zhejiang Lab. His current research interests include the design of high-resolution 4D imaging radar, classification and tracking algorithm for traffic participants, and multi-sensor fusion algorithm for ADAS.
\endbio

\bio{}
\quad \textbf{Sijie Ji} is a research fellow at the Department of Computer Science, The University of Hong Kong. She received her Ph.D degree from Nanyang Technological University, Singapore and her B.E. from the East China Normal University. Dr. Ji’s research interests span the broad area of Artificial Intelligence of Things (AIoT), sensing and mobile computing, cyber-physical systems and edge computing. Dr. Ji has offered various services to the research community including reviewers for INFOCOM, TMC, TON, TWC, TCOM, WCL, CL, ICC, Globecom, VTC and Internet of Things Journal.  She ranked have ranked top3 in multiple academic competitions hosted by the three top computer vision conferences (ICCV, CVPR and ACMMM).
\endbio

\bio{}
\quad \textbf{Zhaohui Yang}  received the Ph.D. degree from Southeast University, Nanjing, China, in 2018. From 2018 to 2020, he was a Post-Doctoral Research Associate at the Center for Telecommunications Research, Department of Informatics, King’s College London, U.K. From 2020 to 2022, he was a Research Fellow at the Department of Electronic and Electrical Engineering, University College London, U.K. He is currently a ZJU Young Professor with the Zhejiang Key Laboratory of Information Processing Communication and Networking, College of Information Science and Electronic Engineering, Zhejiang University, and also a Research Scientist with Zhejiang Laboratory. His research interests include joint communication, sensing, and computation, federated learning, and semantic communication. He received the 2023 IEEE Marconi Prize Paper Award, 2023 IEEE Katherine Johnson Young Author Paper Award, 2023 IEEE ICCCN best paper award, and the first prize in Invention and Entrepreneurship Award of the China Association of Inventions. He was the Co-Chair for international workshops with more than ten times including IEEE ICC, IEEE GLOBECOM, IEEE WCNC, IEEE PIMRC, and IEEE INFOCOM. He is an Associate Editor for the IEEE Communications Letters, IET Communications, and EURASIP Journal on Wireless Communications and Networking. He has served as a Guest Editor for several journals including IEEE Journal on Selected Areas in Communications.
\endbio

\bio{}
\quad \textbf{Zhaoyang Zhang}  received his Ph.D. from Zhejiang University, Hangzhou, China, in 1998, where he is currently a Qiushi Distinguished Professor. His research interests are mainly focused on the fundamental aspects of wireless communications and networking, such as information theory and coding theory, network signal processing and distributed learning, Al-empowered communications and networking, and synergetic sensing, computing and communication, etc. He has co-authored more than 300 peer-reviewed international journal and conference papers, including eight conference best papers awarded by IEEE ICC 2019 and IEEE GlobeCom 2020, etc. He was awarded the National Natural Science Fund for Distinguished Young Scholars by NSFC in 2017.
\endbio

\end{document}